\documentclass[]{aastex62}

\usepackage{amsmath}
\usepackage{stackengine}
 \usepackage{mathtools}

\newcommand{\noop}[1]{}
\newcommand{\okina}{\textquoteleft}

\newcommand{\Ou}{1I/{\okina}Oumuamua}
\newcommand{\borisov}{2I/Borisov}

\newcommand{\be}{\begin{eqnarray}}
\newcommand{\ee}{\end{eqnarray}}

\newcommand{\MSun} {\mbox{$M_{\odot}$}}

\begin{document} 

\title{Oumuamuas passing through molecular clouds}


\author[0000-0002-5003-4714]{Susanne Pfalzner} 
\affiliation{J\"ulich Supercomputing Center, Forschungszentrum J\"ulich, 52428 J\"ulich, Germany}
\affiliation{Max Planck Institute for Radio Astronomy, Auf dem H\"ugel 69, 53121 Bonn, Germany}

\author{Melvyn B. Davies} 
\affiliation{Lund Observatory, Department of Astronomy and Theoretical Physics, 
Box 43, SE-221 00 Lund, Sweden}

\author{Giorgi Kokaia} 
\affiliation{Lund Observatory, Department of Astronomy and Theoretical Physics,
Box 43, SE-221 00 Lund, Sweden}

\author[0000-0003-3257-4490]{Michele T. Bannister}
\affiliation{School of Physical and Chemical Sciences --- Te Kura Mat\={u}, University of Canterbury,
Private Bag 4800, Christchurch 8140,
New Zealand}


\email{s.pfalzner@fz-juelich.de}

\begin{abstract}
 The detections of \Ou\ and \borisov\ within just two years demonstrate impressively that interstellar objects (ISOs) must be common
 in the Milky Way. Once released from their parent system, these ISOs travel for Gyr through interstellar space. While often imagined as empty, interstellar space contains gas and dust most prominent in the form of molecular clouds. Performing numerical simulations, we test how often ISOs cross such molecular clouds. We find that the ISOs pass amazingly often through molecular clouds. In the solar neighbourhood, ISOs typically spend 0.1-0.2\% of their journey inside molecular clouds, for relative slow ISOs ($<$ 5 km/s) this can increase to 1-2\%, equivalent to 10 - 20 Myr per Gyr. Thus the dynamically youngest ISOs spend the longest time in molecular clouds. In other words, molecular clouds must mainly contain relatively young ISOs ($<$ 1-2 Gyr). Thus the half-life of the seeding process by ISOs is substantially shorter than a stellar lifetime. The actual amount of time spent in MCs decreases with distance to the Galactic Centre. We find that ISOs pass so often through MCs that backtracing their path to find their parent star beyond 250 Myr seems beyond the point. Besides, we give a first estimate of the ISO density depending on the galactic centre distance based on the stellar distribution.  


\end{abstract}


\section{Introduction}
\label{sec:intro}

During the planet formation process, enormous numbers of planetesimals are produced functioning as building blocks for planets. However, many of these planetesimals are never incorporated into planets but remain unused. Many of them are ejected from their parent system altogether. This ejection can happen by various mechanisms  \citep[for example,][]{Brasser:2010, Kaib:2011, Hanse:2018, Veras:2014, Do:2018, MoroMartin:2019} that stretch over the entire life of a star, from its formation phase to its eventual end. Once ejected, these planetesimals become interstellar objects (ISOs) that drift through the Galaxy. The vast majority of these icy rocks are long-lived and cruise the Galaxy for billions of years \citep{Guilbert:2015}. 

For a long time, it was unclear how many ISOs are present in the Milky Way. The estimates ranged from as few as 10$^{9}$ pc$^{-3}$ to as many as 10$^{16}$ pc$^{-3}$ \citep[e.g.][]{Whipple:1975, Francis:2005,Engelhardt:2017}. This long-standing argument about the abundance of ISOs was, if not solved, at least considerably constrained, by the recent detections of the 140-m near-inert object \Ou\ \citep{Meech:2017} and the active comet 2I/Borisov \citep{Fitzsimmons:2019, Jewitt:2019}. The efficiency of the detection of \Ou\ during the relevant surveys is the basis for a first observational estimate of the ISO density in interstellar space. The result is an amazingly high ISO density of $\approx$ 10$^{15}$ pc$^{-3}$ ISOs in the local interstellar space \citep{Meech:2017, ISSI}. 

 It is often implicitly expected that these ISOs will have relatively little interaction while travelling through interstellar space remaining in a more or less pristine state. However, even interstellar space is not empty but contains gas and dust mainly concentrated in giant molecular clouds (GMCs). 
GMCs are transient features with a median lifetime of 40 Myr  \citep{Williams:1997, Vallini:2016}. Within GMCs, the gas can condense to form dense cores, and further contraction ultimately may lead to the formation of stars.
The angular momentum of the infalling material leads to the formation
of gaseous disks around these protostars, from which planetary system may form
under favourable conditions. Possibly, ISOs within protoplanetary discs may even help seed planet formation 
 \citep{Pfalzner:2019}. 

Here we will investigate the connection between ISOs and GMCs from two perspectives - that of the ISOs and that of GMCs. When ISOs interact with the dust and gas of the GMC, their dynamics and surface properties might change. The expected effect depends on how often ISOs pass through GMCs and also the relative velocities. These processes have not yet been studied in detail.  Here we determine the relative importance of the interaction of ISOs with GMCs. The prime parameters are the frequency of the passage of ISOs through GMCs and the relative velocity of such encounters. 

The general assumption is that GMCs contain only gas and dust, and eventually the stars forming from them. However, the ISO density in the ISM leads inevitably to the conclusion that GMCs must contain ISOs as an additional component. The question is whether the presence of ISOs plays any role in the context of molecular clouds. The relevance of ISOs depends on their number within GMCs at any given time, which we will determine here. 

We use the simulations performed by \citet{Kokaia:2019} to infer the trajectories of stars and giant molecular clouds (GMCs) in the Galaxy. These trajectories allowed them to determine how often stars pass through GMCs.  For instance, they found that the Sun hits 1.6 $\pm$ 1.3 GMCs per Gyr. These simulations can be generalised to ISOs by employing the fact that the velocity of the ISOs is the velocity of the stars plus the velocity component that led to the ISOs leaving their parent system. In most cases the ISO velocity relative to the star is much smaller than the stellar velocity itself.

We use the existing set of simulations to determine the frequency of ISOs passing through GMCs and the resulting fate of the ISOs. In section 2, the numerical model underlying these simulations is explained. In subsection 3.1 we will present the results and will see that the relative velocities between the ISOs and GMCs are the critical parameter in this context and that the interaction frequency also depends crucially on the location of the GMCs and ISOs in the Galaxy. In subsection 3.2, we determine the number of ISOs typically present in a GMC and the renewal rate of the ISO reservoir. In part 4 we discuss the consequences for our understanding of ISOs.\\

\section{Method}

We follow the orbits of the stars to assess how often ISOs pass through GMCs. Here we describe the Galactic potential used to integrate ISO trajectories with the GMC properties adopted. For further details of the calculations, we refer to \citet{Kokaia:2019}.

\subsection{Simulation set-up}
The orbits of the stars in the Galaxy are determined by integrating their path through a three-component axisymmetric potential $\Phi_G(R; z)$ \citep{Garcia:2001} of the form
\be
\Phi_G(R; z) = \Phi_h(R; z)+ \Phi_b(R; z) + \Phi_d(R; z),
\ee
where the three components $\Phi_h, \Phi_b$ and $\Phi_d$ represent the halo, bulge and disk, respectively. $R$ stands for the Galactocentric cylindrical radius and $z$ the distance to the Galactic plane.  The halo has the form of a spherical Plummer potential \citep{Plummer:1911} and the disk is a Miyamoto disk \citep{Miyamoto:1975}.

The GMCs are characterised by their spatial distribution, mass function, density profile and lifetimes. GMCs are primarily located in the spiral arms of the Galaxy. Here we do not consider the spiral arm gravitational potential explicitly, but instead distribute our model GMCs in the spiral pattern observed by \citet{Hou:2014}. These can be fitted by 
\be
\ln \frac{R}{R_i} = (\theta-\theta_i) \tan \psi_i,
\ee
where $R_i$ is the starting radius, $\theta_i$ the starting angle and $\psi_i$ the pitch angle of GMCs. See Table 2 of \citet{Kokaia:2019} for the values
of $R_i$ and $\psi_i$ used.%

For the mass function of GMCs, we use the fit to observations by \citet{Rosolowsky:2006} given by
\be
\frac{dN}{dM} = (\gamma + 1) \frac{N_0}{M_0} \left( \frac{M}{M_0}\right)^\gamma, M < M_0,
\ee
where $M_0$ is the maximum mass of 3 $\times$ 10$^6$ \MSun\ and the minimum mass is $10^4 \MSun$; $N_0$ = 36 and $\gamma$=1.53 are constants. The GMC mass varies with age as
\be
M(t) = \left[ -0.25 \left(\frac{t-t_0}{10}\right)^2 + \frac{t-t_0}{10}\right] \times M_i,
\ee
where $M_i$ is the mass drawn from equation (3).
The shape of this function is based on simulations from \cite{Krumholz:2006} and \cite{Goldbaum:2011}. The typical lifetime of a GMC is 40 Myr, where the GMC increases in mass for the first 20 Myr, peaks and then decreases for another 20 Myr. The 40 Myr lifetime is consistent with observations from e.g. \cite{Williams:1997} and \cite{Vallini:2016}. We start our simulation with 6,700 randomly generated clouds. This number of clouds gives the correct molecular gas mass (i.e. matches the observations), given the mass function used for the clouds.
Whenever a GMC reaches the end of its life, a new one is initiated in the next time-step, keeping the total number constant.

\begin{figure}[t]
\includegraphics[width=1.0\textwidth]{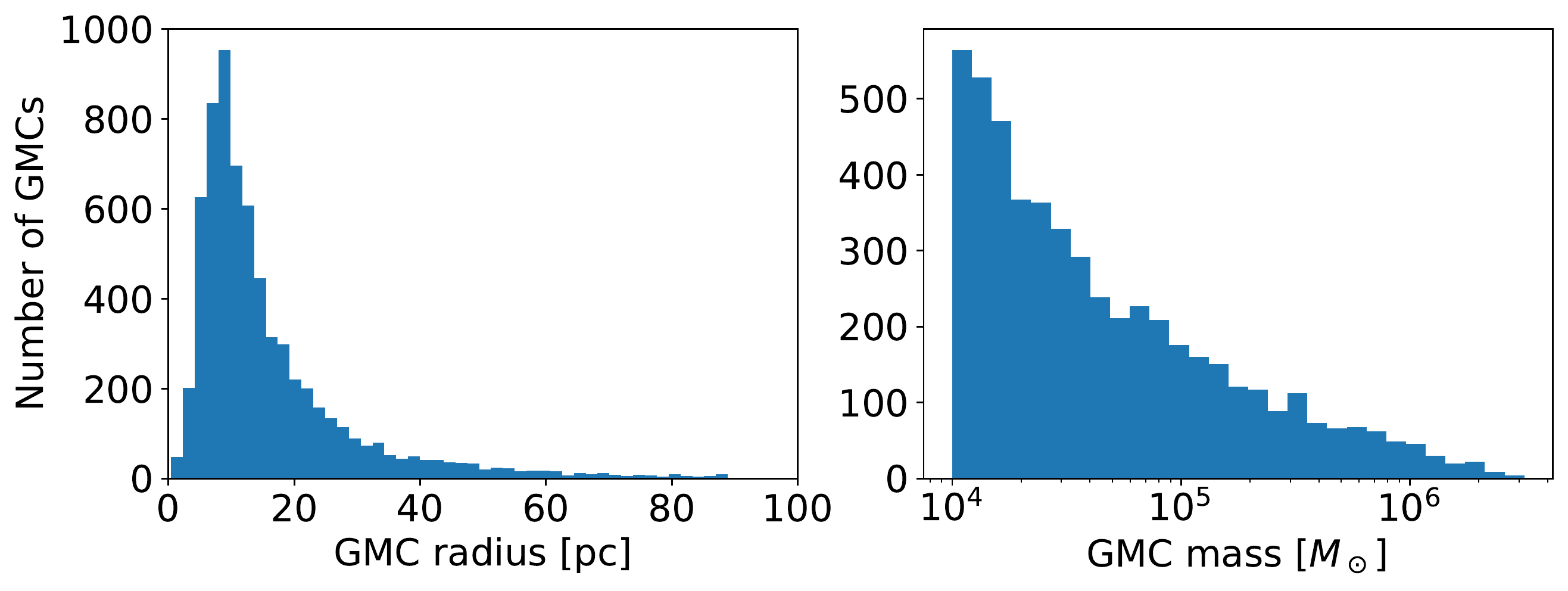}
\caption{Radius and mass distribution of the simulated molecular clouds.}
\label{fig:GMC_radii}
\end{figure}

The H2 gas of the GMC is assigned to one of three categories, following the categorisation by \citet{Roman-Duval:2016}: very dense, dense and diffuse. 
Accordingly, the radii of the GMC is connected to the gas surface densities for these three categories by 

\be 
R_{\rm GMC} = 
\begin{dcases*}
 \frac{1}{229} \left( \frac{M}{1 \MSun} \right)^{1/2.36} 
  &, \text{for 130} \MSun \text{pc}$^{-2} < \rho < $\text{300} \MSun \text{pc}$^{-2}$\\
 \frac{1}{79} \left( \frac{M}{1 \MSun} \right)^{1/2.36} 
  &, \text{for 50} \MSun \text{pc}$^{-2} < \rho < $\text{130} \MSun \text{pc}$^{-2}$\\
  \frac{1}{28} \left( \frac{M}{1 \MSun} \right)^{1/2.36} 
    &, \text{for 25} \MSun \text{pc}$^{-2} < \rho < $\text{50} \MSun \text{pc}$^{-2}$
\end{dcases*}
\ee
where the GMC radius $R_{\rm GMC}$ is in units of pc.
The radial and vertical distributions are provided through H2 observations by \citet{Nakanishi:2006}. Figure \ref{fig:GMC_radii} shows the corresponding distributions of the GMC radii and masses. It can be seen that most GMC have radii in the range of 8 pc $ < R_{\rm GMC} < $ 15 pc.

\newpage
\subsection{Relative velocities between ISOs and parent systems}
\label{sec:prod}

Above we model just the movement of the stars relative to the GMC. However, this can serve as a useful approximation for the relation between the ISOs and the giant molecular clouds. The velocity of the ISOs relative to the GMC, $v_{\tt{ISO}}^{\tt{GMC}}$,  results from two components --- the relative velocity between the star and the molecular cloud $v_{s}^{\tt{GMC}}$ and the velocity between the ISOs and their parent star, $v_s^{\tt{ISO}}$. The latter depends on the type of processes leading to the ejection of the planetesimal
from its stellar host system which turns it into an ISO.  

ISOs are ejected from a star throughout its entire life \citep{Pfalzner:2019}. Judging from the still small sample provided by \Ou\ and \borisov\,, ISOs are typically a few tens to hundreds of meters in size.  In the earliest stages probably a lot of small dust aggregates leave the discs, as indicated by the steep decline of disc mass during the first hundred thousand years \citep{Alexander:2014}. However, unless dust growth is rapid, these particles will mostly be of $mm$ to $cm$ size and therefore much smaller than the ISOs. 

However,  after 2 Myr quite a large number of planetesimals should have grown to sizes of 10 m and above. During that phase, most stars are still part of the stellar cluster they formed in \citep{Kuhn:2019}, therefore close flybys between the cluster stars are relatively common. Such flybys can lead to the truncation of the outer parts of the disc \citep{Vincke:2016, Pfalzner:2018} and as a consequence many planetesimals become unbound, turning into ISOs. The frequency of close flybys depends on the stellar density of the particular group of stars \citep{Vincke:2016, Vincke:2018}. Similarly, the velocity of the created ISOs depends on the actual masses of the involved stars and their periastron distance; typically, they lie in the range 0.5 to 2 km/s (Pfalzner \& Aizpura Vargas Vargas, in preparation).  The velocity of these ISOs relative to the GMCs is then the combination of the stellar velocity and the escape speed and amounts for young stars to approximately 5 km s$^{-1}$ \citep{Hands:2019}. 

While the planets migrate in the still young planetary system, they interact with the debris planetesimal disc. Predominantly, close encounters with the giant planets of the system lead to the ejection of many of a system's remaining planetesimals \citep{Duncan:1987, Charnoz:2003, Raymond:2017}. Mainly planetesimals residing between a few to tens of au from the host star are affected. This type of planetesimal ejection leads to ISOs with typical velocities of 4 to 7 km/s relative to the host star \citep{Adams:2005}.

\begin{figure}[t]
\includegraphics[width=1.0\textwidth]{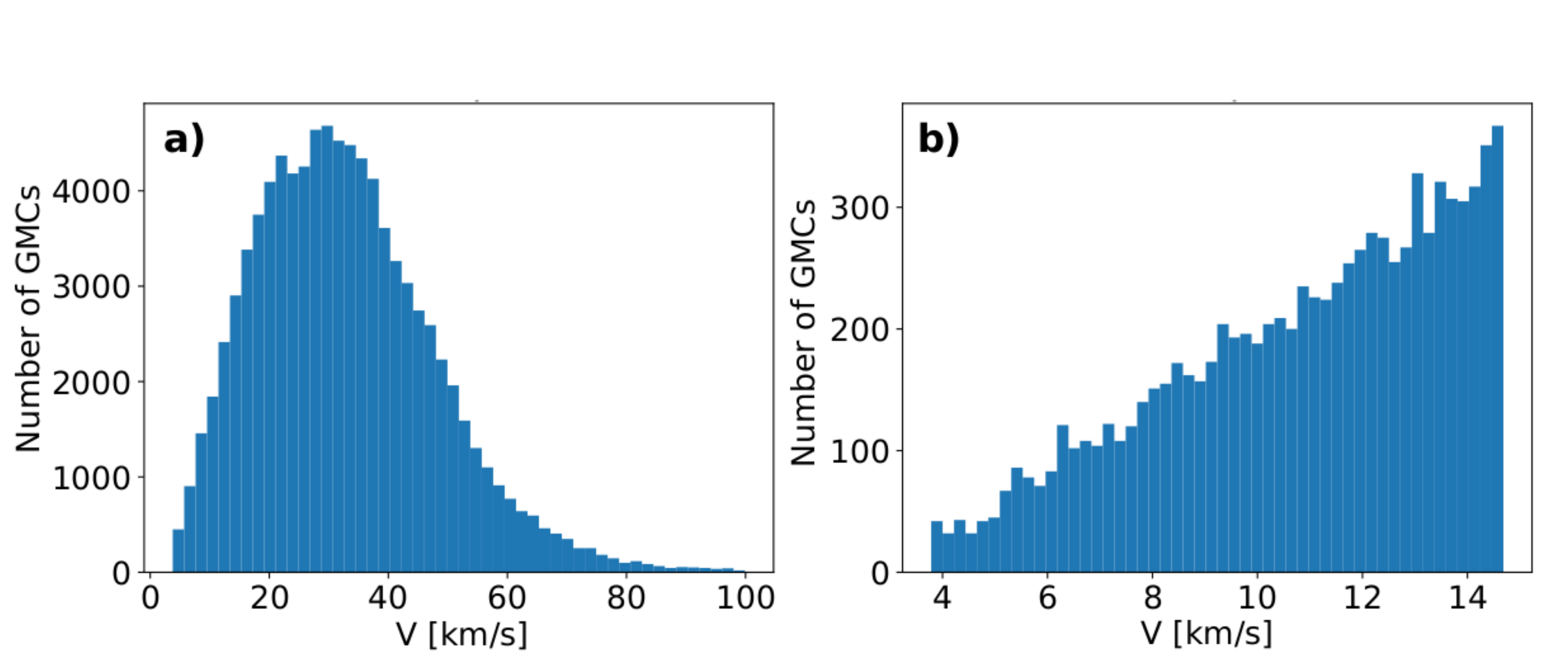}
\caption{Velocity distribution relative to that of the encountered cloud for 1-3 Myr old stars at radial distances between 7.5 and 8.5 kpc from the galactic centre. a) shows the total distribution and b) the ten per cent of slowest stars. }
\label{fig:veldist}
\end{figure}

Afterwards, throughout the star's lifetime, predominately icy planetesimals will be gently lost, as they drift from the distant fringes of the star's Oort cloud under the nudging of the Galactic tide and passing field stars \citep{Brasser:2010, Kaib:2011, Hanse:2018}. This drifting away happens at quite low velocities relative to that of the star ($<$ 0.5 km/s). Finally, the remainder of the system's Oort cloud will be shed to interstellar space once the star leaves the main sequence and loses mass \citep{Veras:2011, Veras:2014, Do:2018, Rafikov:2018, MoroMartin:2019}. The velocity of the so created ISOs depends on the mass of the star but rarely exceeds 0.3 km/s (Veras, private communication).

As mentioned above, it is still unclear which of these processes is the dominant source of ISO production. However, indeed, the ISOs ejection velocities rarely exceed 10 km/s. In most cases, the ejection velocity will be $\ll$ 5 km/s, which means that the velocity of the ISOs differs only slightly from those of the parent stars. Thus the approximation we adopt here -- namely that the ISO velocities within the Galaxy follow the same distribution and evolution as their host stars -- is valid.
Throughout this study we assume that the velocity of the ISOs remains unchanged after ejection. This simplification will be discussed in section 4.

\section{Results}
\label{sec:ISO_results}

\subsection{ISOs perspective}

In the simulations, we followed the trajectories of 400 000 particles to determine how often an ISO passes through a GMC.
Naturally, the frequency of these "passages" depends on the relative velocity between the ISOs and the clouds.  Figure \ref{fig:veldist} a) shows the distribution of these relative velocities for 1-3 Myr old ISOs in the solar neighbourhood (Galactic centre distance, $R_{GC}$, between 7.5 and 8.5 kpc). The relative velocities between stars/ISOs and the GMCs cover a vast parameter space with most ISOs having a relative velocity in the range of 30-40  km/s.  Thus the relative velocity to the GMCs is dominated by that of the stars relative to the GMCs. The ejection speed from the parent stars is of minor importance, being on average less than 10\% of that of the stellar contribution.

\begin{figure}[t]
\includegraphics[width=1.0\textwidth]{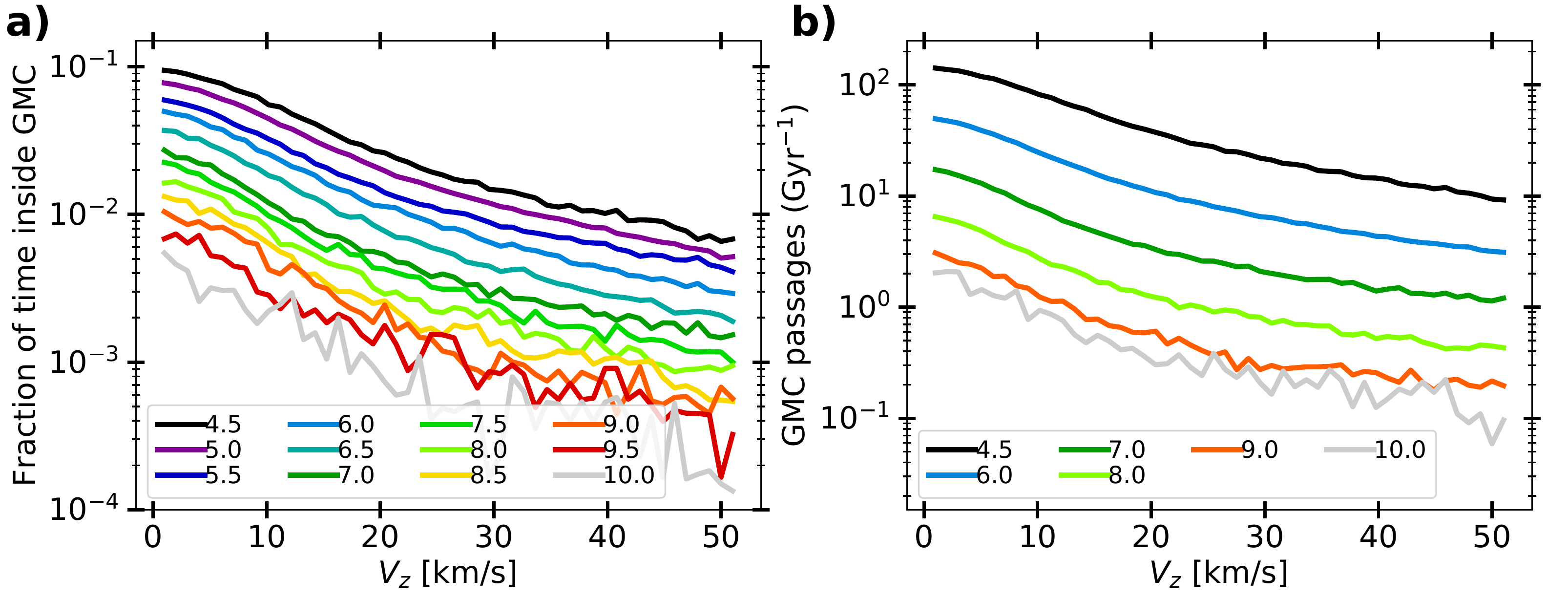}
\caption{The fraction of time spent inside GMCs (a) and the number of GMC passages (b) as a function of ISO velocity. Both values refer to a time period of 1 Gyr. The various colors indicate the results for different distance to the Galactic centre. }
\label{fig:time_velocity}
\end{figure}

ISOs moving with high velocity through the GMC are dynamically less affected than those with low velocities. In turn, fast ISOs are less likely to be captured within a GMC or to be strongly deflected. On the other hand, erosion might be more of a problem for fast ISOs, when small dust particles hit them at high velocities \citep{Housen:2011}. 

Fig. 1b shows a zoom-in into the regime of the low-velocity ISOs. These are the ones that are more likely to be captured. Their number decreases nearly linearly with relative velocity. However, of the 1-3 Myr old ISOs in the solar neighbourhood, about 66\% have a velocity below 10 km/s relative to the GMCs.

The question is how much time do ISOs spend inside molecular clouds. The in-cloud time depends on the cloud properties and individual ISO velocity.  Figure  \ref{fig:time_velocity}a) shows the average fraction of time that ISOs spend inside a GMC as a function of velocity for ISOs in the age range 1-3 Gyr\footnote{More precisely, ISOs ejected from host 1-3 Gyr old host stars}.  Generally, the lower the velocity of the ISO, the longer it spends inside GMCs. The different colours indicate the distance to the Galactic centre.  The light green curve is relevant for the solar neighbourhood ($R \sim$ 8 kpc). In the solar neighbourhood, ISOs with a velocity of $\sim$ 30 km/s spend about 0.2\% of their time inside molecular clouds. The ISOs at the low-velocity end of the distribution, moving with 5km/s, even $\sim$ 1\% of their time inside molecular clouds. In other words,  ISOs in the solar neighbourhood typically spend 10 Myr per Gyr of their interstellar journey travelling through molecular clouds.

 Figure \ref{fig:time_velocity}b) shows that slow ISOs also do pass more often through GMCs than fast ISOs.  In the solar neighbourhood, fast ISO ($v_z \sim$ 50 km/s) pass on average through 0.2 GMCs per Gyr, whereas slow ISOs encounter in the same period typically 4 to 6 GMCs. On first view that seems counter-intuitive.
 The reason why slower ISOs spent more time in GMCs is that all GMCs reside within about 100 pc of the galactic plane. Therefore, ISOs having low velocities in the z-direction are unable to reach above the GMCs as they orbit around the Galaxy; thus they strike them more often; whereas ISOs having higher speeds in the z-direction spend much of their time at larger values of z where there are no GMCs. This leads to ISOs with low values of $v_{\rm z}$ passing through more GMCs compared to ISOs having larger values of $v_{\rm z}$ and contributes to the enhanced time that slow ISOs spend in GMCs.

\begin{figure}[t]
\includegraphics[width=1.0\textwidth]{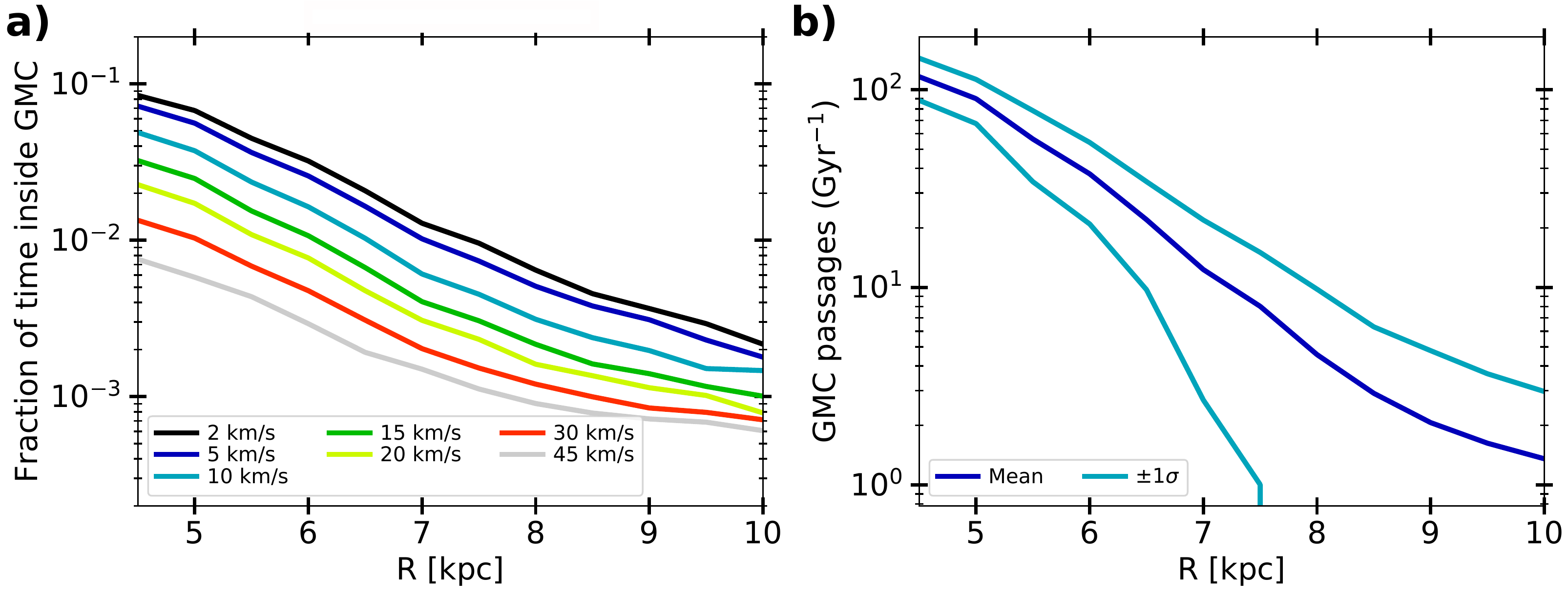}
\caption{(a) The fraction of time an ISO spends inside GMCs and (b) the number of GMC passages per ISO for ISOs initially having a velocity in the z direction of 5 km/s. Both shown as a function of galactic centre distance. Both values are derived for simulations following the trajectories of ISOs over
a time period of 1 Gyr. For figure (a), the various colors indicate the results for different initial ISO velocities in the z direction.}
\label{fig:galactic}
\end{figure}

Already Fig. \ref{fig:time_velocity} showed that the fate of an ISO depends strongly on its location within the Milky Way. Figure \ref{fig:galactic}a shows even more clearly how the time spent inside GMCs depends on the distance to the Galactic centre. This time decreases the further the distance to the Galactic centre. This general trend holds  \footnote{except for the Galactic centre itself} independent of the actual velocity of the ISO. At 4.5 kpc ISOs spent about ten times more time inside GMCs than in the solar neighbourhood, which amounts for slow ISOs (v= 5 km/s) in spending up to an amazing 8\% of their life in GMCs. The main reason is that the GMC density is higher close to the Galactic centre. In summary, many ISOs travel for extended periods unhindered through interstellar space, however, those that move relatively slowly or close to the Galactic centre can spend more time in molecular clouds other than the one in which they were formed.  

Figure \ref{fig:galactic}b) shows that also the number of GMC passages depends strongly on the galactic centre distance.  In the solar neighbourhood, ISOs of velocity 5 km/s pass typically through a handful of GMCs per Gyr, whereas in the same period ISOs at 4.5 kpc pass through 100 GMCs.

For individual ISOs, the time spent inside GMCs and the number of GMC passages can vary considerably. In the solar neighbourhood, the time inside GMCs can differ by orders of magnitude, in particular so for fast ISOs. The relative spread in time spent inside GMCs is much less close to the Galactic centre. Similarly, the number of GMC passages shows a wider relative spread for values in the solar neighbourhood than at 4.5 kpc (see Fig. \ref{fig:galactic}b)). For example, in the solar neighbourhood, the 1$\sigma$-range of the number of GMC passages stretches from none at all to eight passages. By contrast, for the same velocity ISOs at 4.5 kpc the 1$\sigma$-range stretches only from  90-110 passages. At more considerable Galactic distances, there exists a small subset of fast ISOs ($\sim$ 1\%) that never passes through GMCs for 10 Gyrs. However, this is such a small sub-group that these are instead the rare exception rather than the rule.

\begin{figure}[t]
\includegraphics[width=0.8\textwidth]{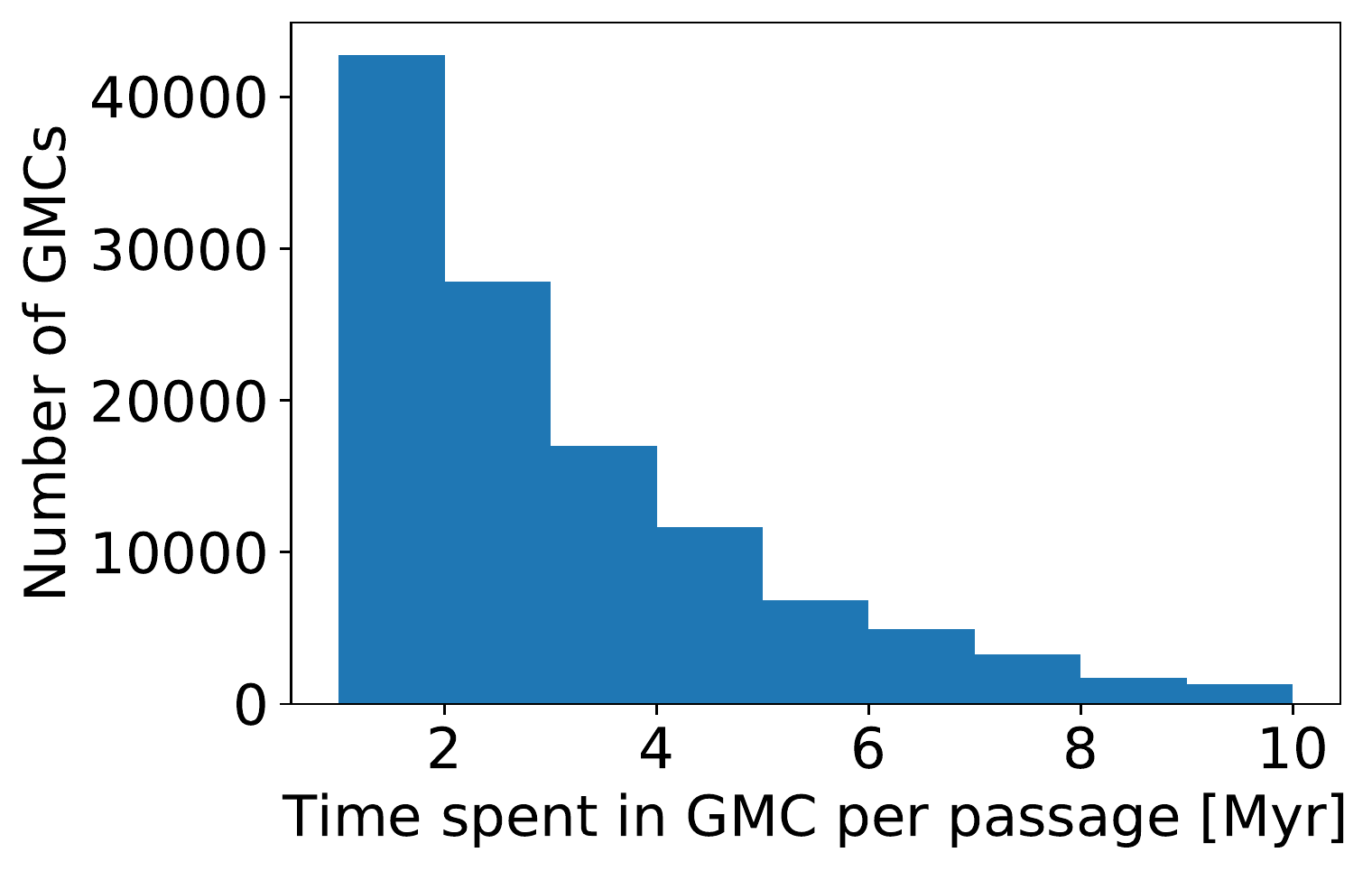}
\caption{Distribution of time spent within GMC per Gyr. }
\label{fig:Time_in_GMC}
\end{figure}

So far, we looked at the total time that an ISO spends inside GMCs. Figure \ref{fig:Time_in_GMC} shows the distribution of the time spent in an individual cloud.  The majority of ISOs spend less than 4 Myr in a given cloud, with the maximum number staying less than 2 Myr. Only a fraction remains in a single cloud for 10 Myr. This is consistent with taking the mean GMCs size (radius  $\sim$ 10pc, size $\sim$ 20 pc) and ISO velocity ($\sim$ 30 km/s $\approx$ 30 pc/Myr) as an approximation, which gives a mean passage time of 0.6 Myr. Thus the mean cloud passage time is much shorter than the lifetime of the cloud, which was assumed to be 40 Myr.  One should note here that ISOs with relatively low relative speeds will be accelerated by the GMC such that all ISOs will pass through GMCs with a minimum speed slightly higher than the escape speed of the GMC (typically around 10km/s).
Thus even the slowest ISOs will pass through an average GMC in $\sim$ 2 Myr
unless they are captured. As a consequence, there is a steady stream of ISOs passing through any individual cloud.\\

\subsection{Cloud perspective}

So far we only considered the fate of the ISOs passing through GMCs. However, from the perspective of the GMC, the large ISO density means that ISOs continuously stream through.  These wandering ISOs are inside the GMCs in addition to those ISOs captured during the cloud formation process itself. The above results imply that the ISOs population passing through a GMC rejuvenates completely approximately every 4 Myr. More specifically, 90\% of ISOs have left by the time the GMC disperses.  Thus the population of ISOs passing through a GMC is replaced  $\gg$10 times during its lifetime. 

The number of ISOs passing through a GMC, $N_{\text{ISO}}^{\text{GM}}$, can be roughly approximated by the volume of the GMC and the ISO density, $\rho_{\text{ISO}}$. It is 
\be
N_{\text{ISO}}^{\text{GMC}} = \frac{4 \pi}{3} R_{\text{GMC}}^3 \times \rho_{\text{ISO}}.
\ee
This means that $\sim$ 10$^{18}$-10$^{19}$ ISOs pass through a typical GMC with $R_{GMC}$ = 10 pc (see Fig. \ref{fig:GMC_radii}). However, this is only a rough estimate; the actual number depends on the size of the cloud, which itself is time-dependent. The cluster radius is proportional to the cluster mass, $ R_{GMC} = C \times M^{1/2.36}$, where $C$ is the same as in Eq. 5 and the GMC mass is in units of solar masses. Although the number of ISO is high their combined mass is quite low. If we assume a density of $\rho$ = 1g/cm$^3$, characteristic of similarly-sized Kuiper belt objects, the combined mass of all ISO present in a 10 pc-sized cloud would correspond to only   
10$^{4} M_\oplus$.

As cluster mass varies with age as given in Eq. 4, the number of ISOs present in a given GMC varies as a function of time accordingly,
\be
N_{\text{ISO}}^{\text{GMC}} & = & \frac{4 \pi}{3}  (C \times M^{1/2.36})^3 \times \rho_{\text{ISO}} = \frac{4 \pi C^3}{3} M^{1.27} \times \rho_{\text{ISO}} = \nonumber \\ 
& = & \frac{4 \pi C^3}{3}    \left[ \left( -0.25 \left(\frac{t-t_0}{10}\right)^2 + \frac{t-t_0}{10}\right) \times M_i \right] ^{1.27} \times \rho_{\text{ISO}} .
\ee

\begin{figure}[t]
\includegraphics[width=0.7\textwidth]{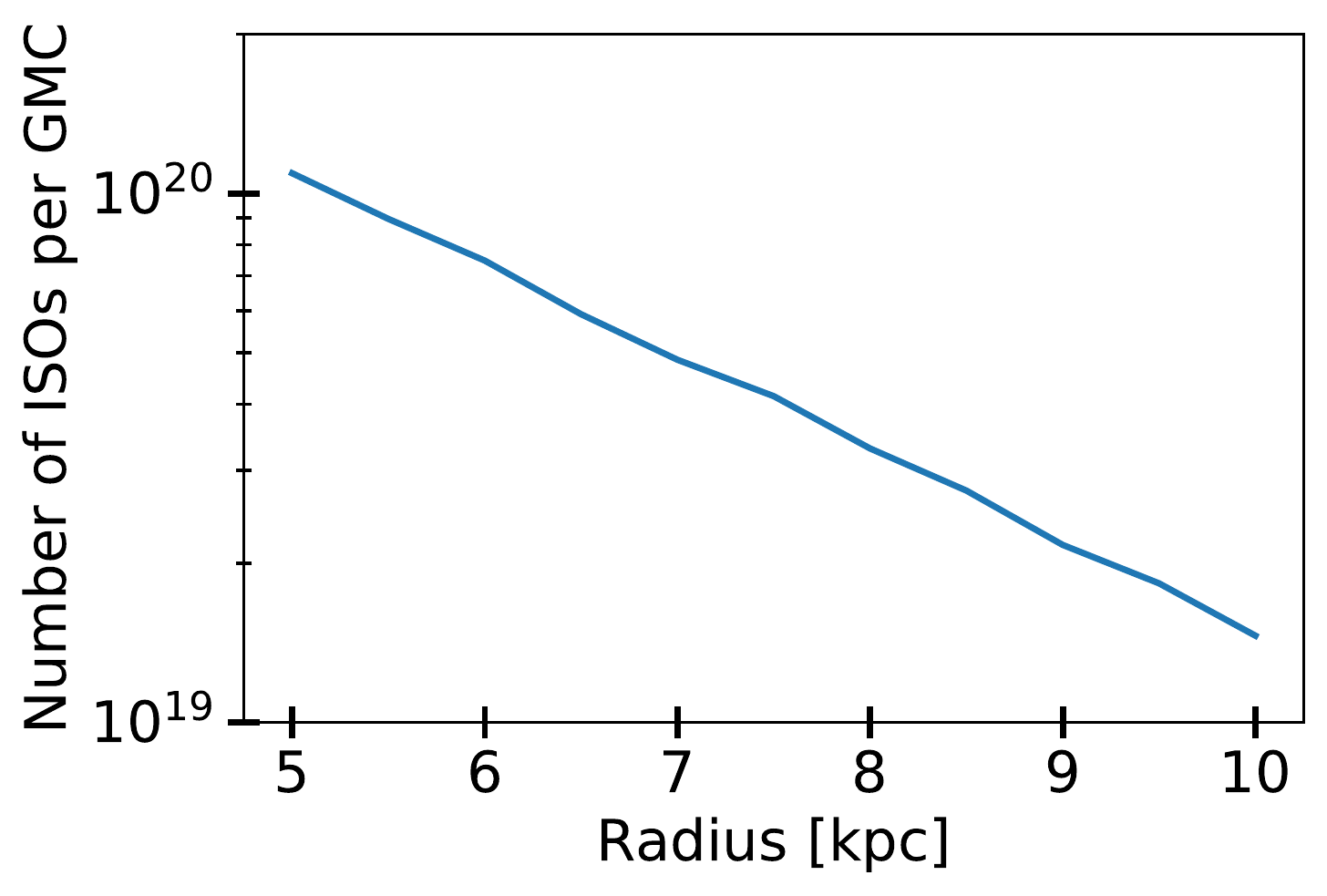}
\caption{Average number of ISOs per cloud as a function of galactic centre distance.}
\label{fig:ISO_in_GMC}
\end{figure}

Figure \ref{fig:ISO_in_GMC} shows the number of ISOs per GMC as a function of galactic centre distance. It can be seen that there are about a factor of ten more ISOs inside GMCs at 5kpc than at 10 kpc. This higher abundance of ISOs means that if ISO presence is in any way significant for GMCs, then GMCs closer to the Galactic centre would be most affected.

%

%
%

\begin{figure}[t]
\includegraphics[width=0.7\textwidth]{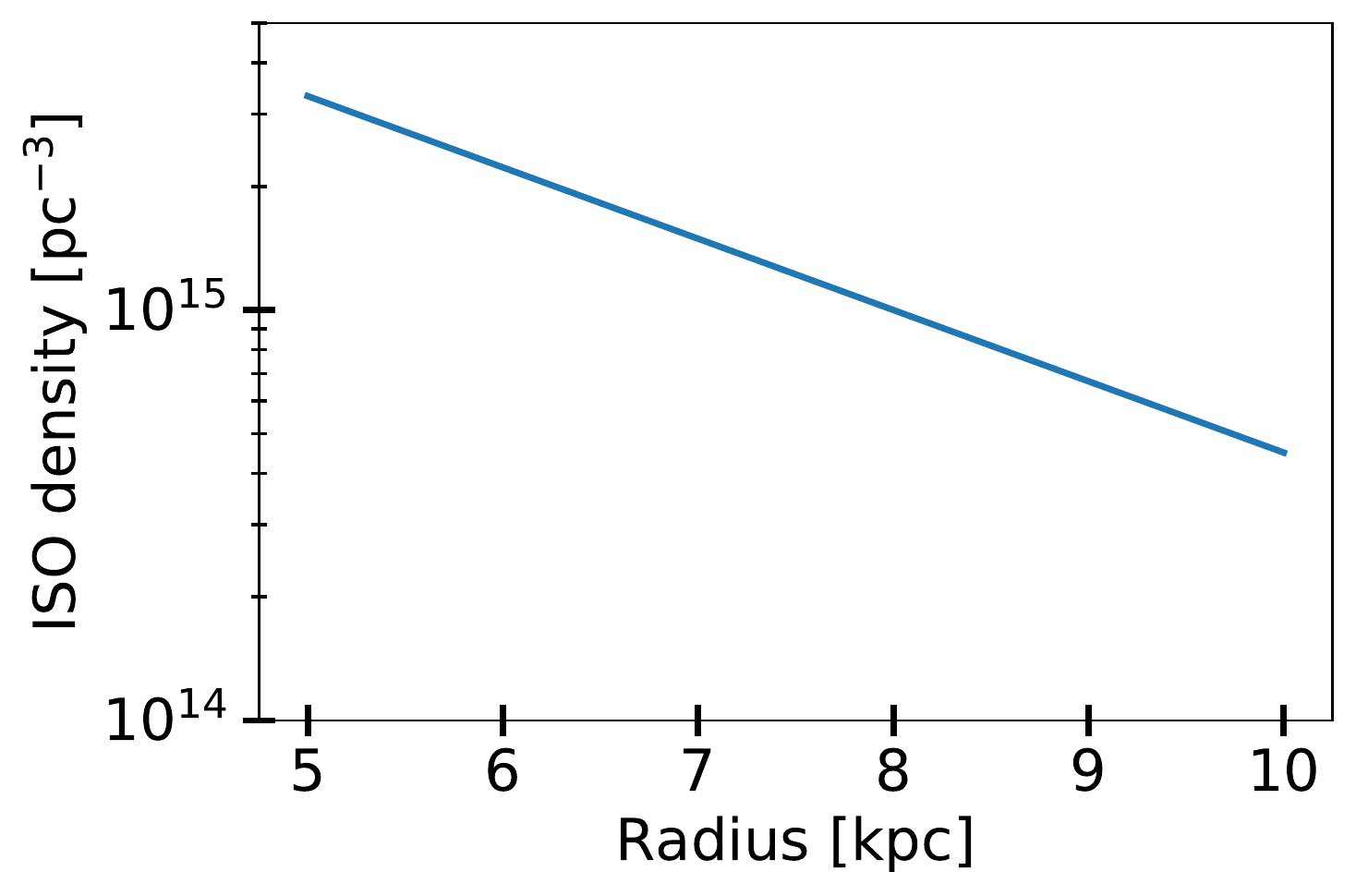}
\caption{ISO density as a function of distance to the Galactic Centre assuming the density follows the stellar surface density.}
\label{fig:ISO_density}
\end{figure}

So far we assumed that the ISO density of the solar neighbourhood, $\rho_{ISO}$ = 10$^{15}$ pc$^{-3}$, is representative for the entire Galaxy. However, it is highly unlikely that the ISO density will be uniform throughout the Galaxy. We saw in section \ref{sec:prod} that the production of ISOs is tightly linked to that of the stars.  Therefore, it seems logical to assume that the ISO production density throughout the Galaxy
follows the stellar distribution at least to some degree. Using the local ISO  density of 10$^{15}$ pc$^{-3}$ as an anchor point,  we used the stellar surface density distribution to give a first estimate of the ISO density within the Galaxy in Fig. \ref{fig:ISO_density}. It shows that the ISO density depends only relatively little on the Galactic centre distance, with the ISO density being only about a factor of three higher at 5 kpc than in the solar neighbourhood. As a consequence the number of ISOs per GMC would depend somewhat more steeply on the galactic centre distance than implied by FIg. 6.

However, one has to be careful with the interpretation of Fig. \ref{fig:ISO_density}. Here the expected production rate is shown, which is not necessarily the same as the present ISO density. 
One can think of many processes that lead to the redistribution of ISOs throughout the Galaxy. For example, ISOs could diffusively spread in a manner similar to that believed to occur for stars due to encounters with spiral arms. On the other hand, ISO production per star could be more efficient closer to the Galactic centre. All these issues will need further study in the future.

\newpage
\section{Discussion}
 
From the results in section \ref{sec:ISO_results}, we conclude that the ISOs in a GMC are mainly those with a relatively low velocity. Some of these might be captured within the GMC, but most will pass unhindered through it.  Predominantly, slow ISOs are produced by stars that have a low velocity relative to GMCs and were released gently from their parent star (i.e. at low ejection speeds). 

Young stars have the lowest velocity dispersion. Therefore, slow ISOs mainly originate from young stars and as a consequence, are young themselves. Consequently, we can conclude that the ISOs present in GMCs are predominately younger than 1-2 Gyr and pass through them at a moderate speed, whereas the fewer old ISOs pass quickly through GMCs.

The fact that the ISOs found in GMCs must have been relatively slow means that these ISOs cannot be produced predominantly during the planet instability process. The reason is that ISO  ejection during the planet instability phase due to scattering processes leads to relatively high velocities (5-10 km/s) compared to other ejection mechanisms (Pfalzner \& Aizpuru Vargas, in preparation.)  By contrast, ISO produced at the end of the lifetime of a star are ejected at low velocity ($\ll$ 1 km/s). However, except for rare high-mass stars, the parent stars are relatively old, which means the stars have a high-velocity dispersion. Therefore, ISOs from the late stages of stars are also unlikely occupants of GMCs. Consequently, ISOs in GMCs originate predominantly from stars that are $\ll$ 1 Gyr old and are either ejected during the cluster phase or drifted gently away from the Exo Oort cloud of the parent star when those were still quite young.   

Another consequence is that young ISOs are the ones most likely to be captured in GMCs. They might even participate in the collapse of parts of the GMC which become unstable.  If ISOs function as seeds of planet formation when they become incorporated into the disc around young stars \citep{Pfalzner:2019}, predominantly the young ISOs function as seeds for new planets. 

By contrast, old ISOs wander through interstellar space with little probability of being captured in a GMC or even just passing through one. It is only these fast old ISOs that are the lonely wanderers which we often imagine all ISOs to be.  However, that older ISOs move at a high velocity through GMCs does not necessarily mean that they are no longer affected by these passages. Their relatively high speed means that they encounter the dust particles at high velocities. These high speed encounters with dust particles might have a more substantial impact on the fast old ISOs, eroding their surface significantly. However, this process will require future detailed study.

Above conclusions are based on assuming that there is a direct correlation between ISOs being young and slow. Throughout our study, we used the host stars' motion to estimate the velocity distribution of ISOs relative to the Galaxy and implicitly assumed that this velocity does not change during the process. However, several processes could alter an ISO's velocity during its journey by interacting with the ISM, stars and molecular clouds. These interactions can either lead to slowing down an ISO or accelerating it. The latter happens predominantly when an ISO passes close to a binary star, as in three-body encounters usually the lightest partner is most likely ejected from the system.  Similar ISOs can become accelerated by they pass the outskirts of a GMC in a slingshot action. The opposite effect can be expected when ISOs pass through the interstellar medium or a GMC itself without being actually captured. In these cases, gravitational drag can potentially slow down ISOs. Similar magnetic fields could potentially slow down ISOs in the ISM \citep{Zhang:2020} and sublimation by collisional heating in GMCs could also influence the ISOs' motion \citep{Hoang:2020}. In most of these process, large ISOs are less effected than small ISOs; here small means ISO sizes smaller than tens to hundreds of meters. Consequently, the velocity distribution ( Fig. 2) is probably a good approximation for large ISOs in general, and the young population of small ISOs might change as they age. It will require further detailed studies that include all these processes to determine the development of the velocity distribution of small, old ISOs to see how this affects, for example, the average number of old ISOs per cloud shown in Fig. 6.

Many GMCs are sites of active star formation which also includes massive stars. Such massive stars are short-lived (less than 50 Myr, see, e.g. Zapartas et al. 2017) and end their lives as core-collapse supernovae. The consecutive explosion usually happens close to where they are born. Therefore, ISOs that pass through a GMC might encounter a nearby supernova that might alter the composition of the ISOs or at least that of their surface layer. As slow, young ISOs spent on average a longer time in GMCs, they are more likely affected by such supernovae explosions than faster ISOs. 

The fate of an ISO does not only depend on its velocity but also its direction. It makes a tremendous difference whether it moves towards or away from the high-density regions closer to the Galactic centre.  As the ejection speed is on average lower than the velocity dispersion of the parent stars, this is only to a small degree determined by the ejection direction itself but mostly by the direction of the velocity of the parent star at the moment of ISO release.

During the last three years, there has been some effort to pinpoint the origin of \Ou\ and  \borisov.  The preferred method is to search trajectories for close passages to stars in the past. There is an ongoing discussion whether this is possible given the uncertainties in stellar positions and velocities \citep{Dyb:2018,Zhang:2018,Feng:2018,Bailer-Jones:2018,PortegiesZwart:2018,Dybczynski:2019,Higuchi:2019,Bailer:2020,Hallat:2020}.
The fact that low-velocity ISOs interact most often and intensely with GMCs also has consequences for tracing ISOs back to their parent star. In the solar neighbourhood slow ISOs pass on average through 4-6 GMCs per Gyr. During these passages, ISOs interact gravitationally with the GMC. Consequently, their path will alter, even if this change is only modest in many cases. However, the consequence is that on average, after only 0.25 Gyr, the path of an ISO is already no longer traceable. Not only that, but the GMC that changed the path of the ISO will have dissolved without a trace after just 40 Myr, making it also impossible to detect the possible points where the ISO deviated from its original root.

We looked at the distribution of ISOs as function Galactic centre distance. However, this is only one aspect when considering variations in ISO density. The actual situation is more complicated. As illustrated in Section \ref{sec:intro}, the ISO production process is strongly linked to the star and planet formation process. Therefore, any variation in the star formation efficiency translates automatically into
a difference in ISO production. 

The ISO density probably also changed over time. As ISOs can likely survive for many Gyr while travelling through interstellar space, their absolute number should have increased with time \citep{Pfalzner:2019}.
Or in other words, the results above can only be applied for the last 2-3 Gyr, and the number of ISOs per GMC was probably considerably lower in the distant past. 

In our simulations, collisions between GMCs and their subsequent mergers were not considered. However, mergers might be relevant in the central areas of the Galaxy due to the higher GMC density. In this area, a more realistic treatment of the GMC dynamics might be required.

 
\newpage 
\section{Summary and conclusions}

With locally 10$^{15}$ objects per cubic parsec, ISOs are quite abundant in interstellar space, so are giant molecular clouds. Naturally leading to the question of how often ISOs pass through clouds and what are the potential consequences of these passages for ISOs and molecular clouds alike. Here we performed numerical simulations to obtain a first impression on the potential consequences of such encounters. In summary, our results show that:

\begin{itemize}
    \item The frequency of ISOs passing through GMCs is a strong function of both the ISO velocity and distance to the galactic centre.
    \item In the solar neighbourhood, typical ISOs ($ v \sim$ 30 km/s) spend approximately  0.1-0.2\% of their time inside GMCs. In other words, these ISOs spend approximately 1-2 Myr transversing GMCs for every Gyr they travel through "empty" interstellar space. 
    \item Slow ISOs ($\sim$ 5 km/s) spend on average a larger fraction of time within a GMC; in the solar neighbourhood they are typically located inside GMCs about 1-2\% of the time (equivalent to 10-20 Myr per Gyr). The reason is a combination of longer GMC transversal times and a higher frequency of encountering GMCs.
    \item Despite these general trends the time inside GMCs can vary for individual ISOs by many orders of magnitude in the solar neighbourhood.
    \item ISOs close to the Galactic Centre  also spend a larger fraction of their life inside GMCs. Typical ISOs at 4.5 kpc spend nearly ten times more time inside GMCs than in the solar neighbourhood. Here slow ISOs (v= 5 km/s) spend nearly up to 10\% of their life in GMCs. 
    \item The reservoir of ISOs inside GMCs is rejuvenated continuously, with the entire ISO population being replaced about ten times within the lifetime of a GMC. 
\end{itemize}

The conclusion is that ISOs pass surprisingly often through GMCs. This might have several consequences for ISOs and GMCs alike.

When ISOs pass through GMCs, they can be gravitationally and structurally affected. These effect have not been taken into account in this study. However,
the gravitational effect of the GMCs can alter the ISO velocity and in extreme cases, lead to their capture in the cloud. Here the slow and therefore probably preferentially young ISOs are mostly affected. Even if not captured, they change their velocity, which often means also a change in direction. Consequently, tracing back the origin of ISOs is not only hampered by the scattering by stars but more substantial by being scattered by the GMCs they inevitably pass through.

Assuming that most ISO's do not experience a large change of their velocity during their travel, GMCs will mainly contain relatively young ISOs ($<$ 1 Gyr). Thus the recycling process of ISOs is relatively fast. Therefore, it is mostly the relatively young ISOs that might seed the next generation of planets.

ISOs can also be effected on a structural level when passing through GMCs. They face a high collision rate with the dust particles present in the cloud. The actual outcome of this exposure likely depends on the exact composition of the ISOs and the dust; however, high incident rates will lead at least to the erosion of the ISO surface. In extreme cases, this can also lead to ISO restructuring or even destruction.

The above results affect our view of {\Ou} and \borisov. The age of {\Ou}  is not known. However, with a random velocity of $\sim$ 9 km/s from the local standard of rest, \Ou's random velocity is much smaller than that of nearby stars ($\sim$ 50 km/s) \citep{Anguiano:2018}. This velocity indicates that {\Ou} is dynamically young with a statistically derived dynamical age of $<$ 2 Gyr \citep{ISSI}. If we assume {\Ou} to have an age in the range 1-2 Gyr and also take \Ou's low relative velocity into account, {\Ou} could have transversed up to 4-10 GMCs spending 5 to 20 Myr of its journey inside GMCs. Therefore back tracking Oumuamua's origin beyond distances that correspond to 250 Myr travel time is probably not well motivated.

So far, the differences seen between Oumuamua and Borisov are attributed to them having formed differently. However, the above results also open the new option that these differences might be due to a different history during their passage through interstellar space. In this picture, Borisov would have had a comparatively uneventful journey passing through few if any GMCs, arriving here more or less unaltered leading to its appearance as a comet-like object. By contrast, Oumuamua would have encountered many GMCs and undergone major restructuring before reaching the solar system in a somewhat battered state.
However, the effect of GMCs on ISOs needs further investigations to see how likely this scenario is.\\

\acknowledgments
M.T.B. and S.P. thank the International Space Science Institute (ISSI Bern), which triggered this collaboration, and our enthusiastic colleagues on the ISSI \Ou\ team\footnote{\url{http://www.issibern.ch/teams/1ioumuamua/}} for two enjoyable workshops that helped spark this work. M.B.D. acknowledges the support of the Knut and Alice Wallenberg Foundation through grant 2014.0017 ``IMPACT".

\bibliographystyle{aasjournal}
\bibliography{references}





\end{document}